*Correspondence and requests for materials should be addressed to
C.Q. (email: cyqiu@whu.edu.cn) or Z.L. (email: zyliu@whu.edu.cn).

# Large-Angle Bending Transport of Microparticles by Acoustic Half-Bessel Beams

Yixiang Li,[1] Chunyin Qiu,[1]* Shengjun Xu,[1] Manzhu Ke,[1] and Zhengyou Liu[1,2]*

[1]Key Laboratory of Artificial Micro- and Nano-structures of Ministry of Education and School of Physics and Technology, Wuhan University, Wuhan 430072, China

[2]Institute for Advanced Studies, Wuhan University, Wuhan 430072, China

**Conventional microparticle transports by light or sound are realized along a straight line. Recently, this limit has been overcome in optics as the growing up of the self-accelerating Airy beams, which are featured by many peculiar properties, e.g., bending propagation, diffraction-free and self-healing. However, the bending angles of Airy beams are rather small since they are only paraxial solutions of the two-dimensional (2D) Helmholtz equation. Here we propose a novel micromanipulation by using acoustic Half-Bessel beams, which are strict solutions of the 2D Helmholtz equation. Compared with that achieved by Airy beams, the bending angle of the particle trajectory attained here is much steeper (exceeding 90º). The large-angle bending transport of microparticles, which is robust to complex scattering environment, enables a wide range of applications from the colloidal to biological sciences.**

In the past decades, great efforts have been devoted to manipulate microparticles and even living cells under contactless conditions. Many different techniques have been proposed, such as optical tweezers[1-5] and dielectrophoresis.[6-9] Acoustic waves can also exert forces on illuminated objects by exchanging momentum between the objects and sound field.[10-17] In contrast to the other contactless manipulations, the acoustic radiation forces (ARFs) necessitate much less power and thus exhibit weaker damage to objects. This advantage enables the acoustic manipulation with wide applications, e.g., the DNA transfection and targeted drug delivery. Recent studies have also stated that, remarkable interactions can be induced by acoustic waves in multi-body systems,[18,19] through the aid of resonant coupling between artificial structures. Besides the intrinsic properties of the objects, i.e., the geometry and acoustic properties, the ARF depends greatly on the property of the external sound source. Recently, the ARFs produced by nondiffracting acoustic beams (which preserve shape in propagation) have attracted much attention due to the capability of stable manipulation in a long distance. A representative example can be referred to the acoustic pulling effects by utilizing Bessel beams[20-22] or crossed plane waves.[23,24]

Conventionally, it is believed that in homogenous space shape-invariant wave beams travel along a straight line only. This cognition was broken up by the observation of Airy beams in optics:[25,26] the main lobes of Airy beams propagate along a parabolic trajectory. The beams with lateral shift, called self-accelerating beams, have two peculiar features: diffraction-free while propagating and self-healing even if blocked by obstacles. These properties endow the self-accelerating beams with new degree of freedoms in contactless manipulations, e.g., guiding microparticles along a curved trajectory.[27,28] It is worth pointing out that, the Airy beam is only a solution of the 2D Helmholtz equation in the paraxial limit. This condition fails and diffraction occurs when the Airy beam travels along the parabola trajectory and eventually bends into a larger angle. This leads to a serious limitation in practical applications. Recently, a new family of self-accelerating optic beams, i.e., Half-Bessel (HB),[29-32] Mathieu[33-35] and Weber[33,35] beams have been demonstrated theoretically and experimentally. These novel beams are exact solutions of the 2D Helmholtz equation, which preserve their shapes while propagating along circular, elliptic or parabolic trajectories. Therefore, compared to the Airy beam, these nonparaxial accelerating beams can be used to achieve larger bending angles over a longer distance. This is greatly beneficial for manipulating microparticles without hampering by obstacles.[36]

In this paper, we study the ARF acting on spherical microparticles illuminated by an acoustic HB beam. Distinct manipulation behaviors have been observed for the particles with different sizes and acoustic parameters. The phenomena can be explained qualitatively from the competing effect between the gradient force and the scattering force. It is of particular interest that in some situations the particle can be attracted into the strong field region and guided stably along a circularly curved orbit. The bending angle of the transport trajectory is considerably large (exceeding 90°), which is unattainable by using conventional sound beams. The large-angle transport could be greatly useful in the contactless manipulation of microparticles, such as to deliver biomedicines under complex conditions.

**Results**

**Brief introduction for acoustic HB beams.** Similar to its optic counterpart,[29-32] the sound field of the HB beam varies only in the 2D space, e.g. $x$-$y$ plane here. It can be written in term of the velocity potential $\psi_{HB}(x,y)$, i.e.

$$\psi_{HB}(x,y) = \psi_0 \int_0^\pi e^{i\alpha\tau} e^{ik[x\cos\tau + y\sin\tau]} \mathrm{d}\tau \quad , \tag{1}$$

where $\psi_0$ is a unitary velocity potential, $\alpha$ is an arbitrary real number, and $\mathbf{k} = (k\cos\tau, k\sin\tau)$ represents the 2D wavevector in the background fluid. In Eq. (1) the integral from 0 to $\pi$ means that the HB beam consists of a series of plane waves with positive $y$-component of wavevectors, which is crucially different from the conventional Bessel beam, integrating from 0 to $2\pi$.

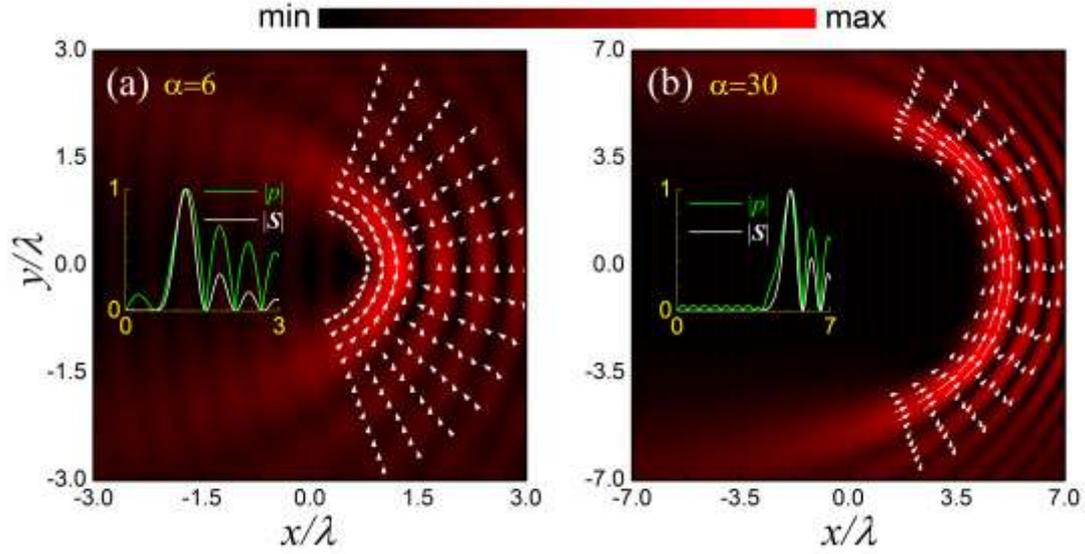

**Figure 1 | Characteristics of the acoustic HB beams.** The amplitude of the pressure field $|p|$ (color) and the corresponding Poynting vector $\mathbf{S}$ (arrow) for the acoustic HB beams with (a) $\alpha = 6$ and (b) $\alpha = 30$, where the length and orientation of the arrow indicate the magnitude and direction of $\mathbf{S}$, respectively. The insets exhibit $|p|$ and $|\mathbf{S}|$ evaluated at $y = 0$, normalized by the corresponding maxima.

To qualitatively show the property of the acoustic HB beam, in Figs. 1(a) and 1(b) we present the pressure field distributions for the cases with $\alpha = 6$ and $\alpha = 30$, together with the corresponding Poynting vector $\mathbf{S} = p\mathbf{u}^*/2$ for comparison. Here $p = -ic_0\rho_0 k\psi_{HB}$ and $\mathbf{u} = -\nabla\psi_{HB}$ stand for, respectively, the first-order pressure and velocity fields, with $\rho_0$ and $c_0$ being the static mass density and sound speed of the surrounding fluid (i.e., water throughout the paper). As shown in Figs. 1(a) and 1(b), both cases exhibit a series of alternate ring-like stripes. Different from the spatial structure of the conventional Bessel beam, each pressure lobe of the HB beam does not close itself, associated with a big bending angle (exceeding 90°) between the head and tail of the beam. Another important feature of the HB beam is that the position dependent Poynting vector always points to the counterclockwise direction, as a direct consequence of the integral from $0$ to $\pi$ in Eq. (1). Besides, the magnitude of Poynting vector oscillates and decays in the radial direction, which is in accordance with the oscillatory pattern of the pressure amplitude. This can be seen more clearly in the insets, plotted for the normalized amplitudes of the pressure and Poynting vector at $y = 0$.

A significant property of the HB beam is its shape-preserving when propagating along the circularly curved orbit. Quantitatively, the HB beam with a larger $\alpha$ gives more accurate nondiffracting propagation over a longer distance, associated with a smaller field gradient along the angular direction. Qualitatively, however, the property of the HB beam does not change much with $\alpha$. Below we focus on the case of $\alpha = 6$, which provides a better visual effect of the ARF distribution (due to the narrower field region involved).

**Evaluation of the ARF exerting on a spherical particle.** For a fluid-surrounded spherical particle, the ARF can be evaluated by integrating the time-averaged Brillouin radiation stress tensor after solving the self-consistent sound field of the scattering problem (see method). Considering the translation invariance along the *z*-axis, the ARF has only in-plane components. For the convenience of presentation, hereafter the ARF is scaled by the unit $E_0 S_c$, where $E_0 = \frac{1}{2}\rho_0 k^2 \psi_0^2$ represents a characteristic energy density, and $S_c = \pi R^2$ is the cross-sectional area of the particle.

**Typical particle manipulations in acoustic HB beams.** In most practical applications, the frequencies concerned are typically in the upper kHz to the lower MHz range, associated with wavelength much larger than microparticles. For simplicity, below liquid droplets are exemplified to demonstrate typical micromanipulations by acoustic HB beams, which essentially exist in elastic particle systems as well. All material parameters involved are listed in table. 1, i.e., the mass density $\rho$, compression modulus $\kappa$ and sound speed $c = \sqrt{\kappa/\rho}$.

| material | $\rho$(kg/m$^3$) | $\kappa$(N/m) | $c$(km/s) |
|---|---|---|---|
| water | 1.0X10$^3$ | 2.22X10$^9$ | 1.49 |
| hexane | 0.66X10$^3$ | 1.14X10$^9$ | 1.07 |
| glycerin | 1.3X10$^3$ | 4.79X10$^9$ | 1.92 |
| air | 1.29 | 1.49X10$^5$ | 0.34 |

**Table 1.** Material parameters involved in the text.

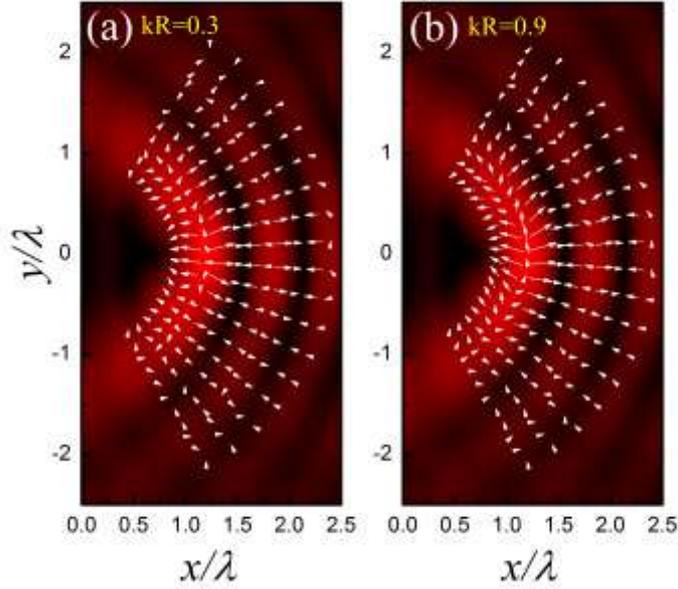

**Figure 2 | Position dependent ARF exerting on a water-surrounded glycerin droplet.** Here the length and direction of the arrow represent the magnitude and orientation of the ARF, and the color displays the amplitude of pressure field copied from Fig. 1(a). The maximal magnitudes of the dimensionless ARFs for the two cases are 0.09 and 0.23, respectively.

The first example considered is a water-surrounded glycerin droplet. Comparing to water, glycerin is harder to be compressed. In Fig. 2 we present the ARF (arrow) distribution for a glycerin droplet located in the acoustic HB beam with $\alpha = 6$, together with the pressure field (color) for comparison. Two different droplet sizes are considered, i.e., $kR = 0.3$ and $0.9$. For the case of $kR = 0.3$, Fig. 2(a) shows that the angular component of ARF is much weaker than its radial component, such that the droplet prefers to stay in the dark stripes. For the case of $kR = 0.9$, the situation is somehow different. The ARF distribution in Fig. 2(b) shows a remarkable growth of the angular component in the brightest stripe (i.e., main lobe), which suggests that a glycerin droplet located there can move along the curved trajectory. However, the bending transport is unstable since the droplet will be simultaneously attracted into the neighboring dark stripe, where the angular ARF is negligible again.

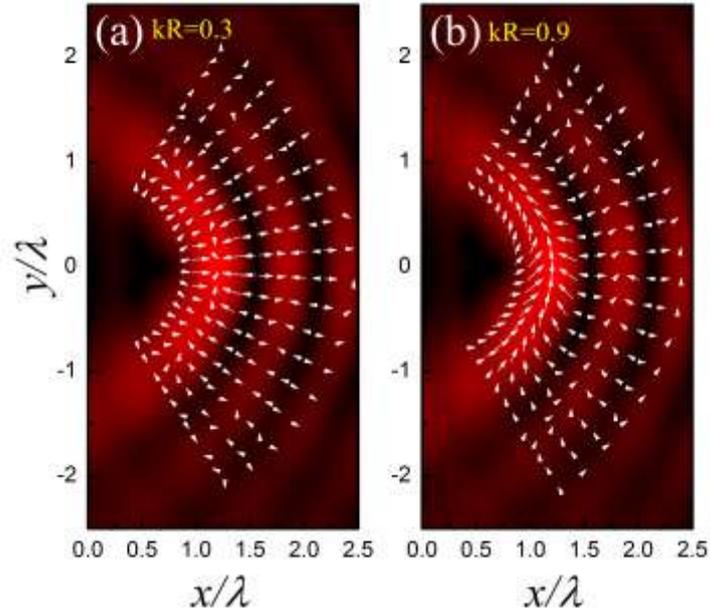

**Figure 3 | Position dependent ARF exerting on a hexane droplet.** The maximal magnitudes of the dimensionless ARFs for the two cases are 0.23 and 2.36, respectively.

The second example under consideration is a water-surrounded hexane droplet. In contrast to the first example, now hexane is relatively easier to be compressed than water. Again, $kR = 0.3$ and $0.9$ are considered, associated with the numerical ARF distributions provided in Figs. 3(a) and 3(b), respectively. Different from the glycerin droplet case, the hexane droplet prefers to stay at the bright stripes. As $kR$ increases, as shown in Fig. 3(b), the angular ARF grows and the droplet tends to migrate along the bright stripes. The direction of the droplet motion is consistent with the orientation of the Poynting vector indicated before, i.e., in a counterclockwise manner. Particularly, in contrast to the glycerin case, now the bending transport is stable since the droplet is always attracted to the bright stripes carrying with considerable angular force components, especially along the main lobe. The bending angle is very large, which exceeds $90^o$ in this case ($\alpha = 6$).

**Qualitative interpretations.** The above phenomena can be understood as follows. It is well-known that for a incident sound field with non-negligible spatial gradient, the ARF acting on a Rayleigh particle (associated with weak sound scattering) can be estimated by the simple formula,[37,38]

$$\mathbf{F} \approx -\pi R^3 \nabla \left( \frac{f_1}{3\kappa_0} |p|^2 - \frac{f_2 \rho_0}{2} |\mathbf{u}|^2 \right). \quad (2)$$

Here the coefficients $f_1 = (\kappa_s - \kappa_0)/\kappa_s$ and $f_2 = 2(\rho_s - \rho_0)/(2\rho_s + \rho_0)$, which are closely related to the parameter contrast between the scatterer (denoted by *s*) and the background fluid (denoted by 0). This expression reveals a competition between the gradients of the pressure and velocity intensities. For the specific HB beam, in Fig. 4 we present a phase diagram for stability analysis, where the green curve indicates a critical situation that separates the preferences to the strong and weak (pressure) field regions. Roughly speaking, a soft droplet (e.g., hexane) tends to be attracted to the strong field region, whereas a hard droplet (e.g., glycerin) prefers to stay at the weak field region.

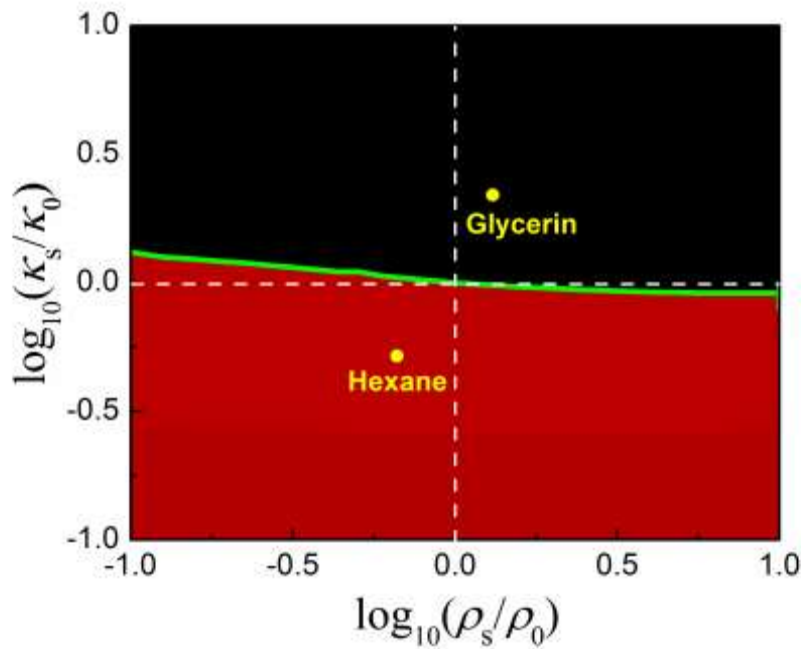

**Figure 4 | Phase diagram for stability analysis.** The microparticles with acoustic

parameters above and below the green curve tend to stay at the dark and bright field regions, respectively.

The above interpretation is mainly responsible for the radial direction of the HB beam that carries with remarkable spatial gradient of sound field. For the angular direction, the spatial gradient is much weaker, especially for the HB beam with large $\alpha$. Remember that for a plane wave without any gradient, the ARF acting on a Rayleigh particle can be estimated by the analytical formula[37,38]

$$\mathbf{F} \approx \frac{4\pi \mathbf{S}}{9c_0} k^4 R^6 \left( f_1^2 + f_1 f_2 + \frac{3}{4} f_2^2 \right), \tag{3}$$

where $f_1$ and $f_2$ have been defined above. The ARF in Eq. (3) stems from the scattering of the particle and points to the direction of the Poynting vector $\mathbf{S}$. This scattering force, proportional to $R^6$, is much more sensitive to the particle size than that induced by field gradient, which is proportional to $R^3$ as indicated in Eq. (2). Therefore, if the particle is small enough, the gradient force (even in the angular direction) could be dominant over the scattering force. This explains the fact that a hexane droplet of $kR = 0.3$ positioned at the main lobe will be attracted into the brightest spot, as shown in Fig. 3(a). As the particle size increases, the scattering force takes effect gradually and even takes a dominant role. This explains the fact for $kR = 0.9$, the hexane droplet tends to move along the bright circular stripes, accompanying with a counterclockwise direction which is consistent with that of Poynting vector (see Fig. 1). It is also predictable that for the beam with larger $\alpha$, the stable bending transport of the hexane droplet can be realized for a much smaller size due to the weaker angular gradient of sound field, together with an increased transport path and a steeper bending angle (approaching 180°). For the glycerin droplet with $kR = 0.9$, the scattering force is also remarkable at the brightest stripe [see Fig. 2(b)]. However, the angular transport is unstable since the droplet will be attracted simultaneously toward the neighboring dark field regions, which carry with small Poynting vectors and thus weak angular scattering force. Therefore, the state of the glycerin droplet is rather complicate, which depends on the specific initial state and dynamic process.

Note that the above analysis is also applicable for elastic microparticles, considering the fact that the descriptions of the gradient and scattering forces in Eq. (2) and Eq. (3) hold for elastic particles as well, except that the compression modulus $\kappa_s$ in the coefficient $f_1$ is replaced by the bulk modulus of the solid, i.e., $B_s = \kappa_s + 2\mu_s/3$, with $\mu_s$ being the shear modulus of the elastic particle.

**Switchable manipulations on air bubbles.** So far we have discussed two kinds of droplets with sound parameters comparable with water. Now we study an often-encountered system with strong parameter contrast, i.e., an air bubble immersed in water.

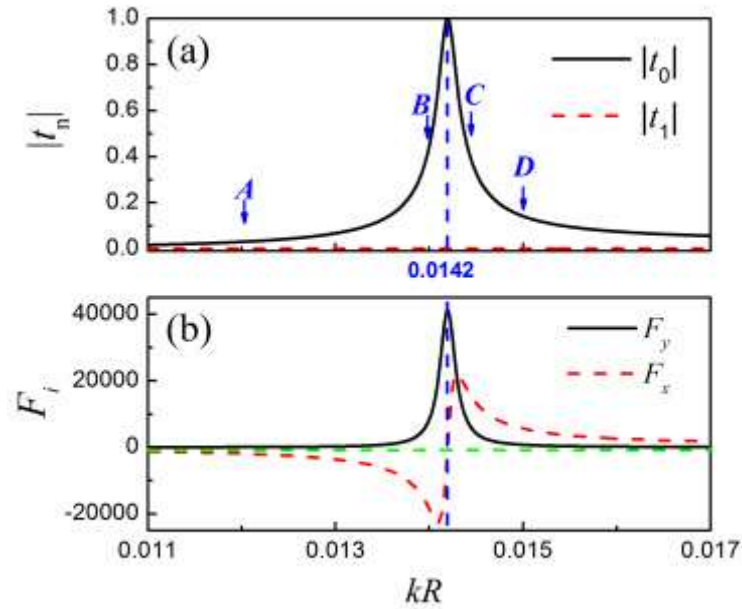

**Figure 5 | Switchable manipulations on air bubbles.** (a) The lowest two Mie scattering coefficients for the air bubble immersed in water, plotted as a function of $kR$. (b) The dimensionless ARF components evaluated for the air bubble located at the position $(1.38\lambda, 0)$.

In this system, it is well-known that the scattering effect can be very strong, even for the bubble with deep subwavelength size ($kR \ll 1$). This can be seen from the

**10 / 20**

amplitude of the Mie coefficient $t_0$ in Fig. 5(a), which exhibits a strong monopole resonance near $kR = 0.0142$. [Note that in this deep subwavelength region, the scattering contributed from the other angular channels is negligibly small, e.g., the dipole channel $t_1$ shown in Fig. 5(a).] It is anticipated that the strong monopole scattering can remarkably enhance the angular component of the ARF. To illustrated this, in Fig. 5(b) we present the dimensionless y-component of the ARF $F_y$ (black solid line) varying with $kR$, evaluated for the bubble located at $(1.38\lambda, 0)$, a position slightly deviating from the brightest spot of the main lobe, i.e. $(1.20\lambda, 0)$. It is exhibited that the dimensionless $F_y$, i.e., the angular ARF at this bubble position, is as high as ~40000 at resonance, associated with an enhancement of $\sim 10^4$ times with respect to the glycerin and hexane cases. In Fig. 5(b) we also provide the corresponding ARF in the radial direction, i.e., $F_x$ (red dashed line). It suggests a switchable manipulation when crossing the monopole resonance: for the case before resonance, $F_x$ is negative and the bubble will be dragged toward the brightest spot; for the case after resonance, $F_x$ is positive and the bubble will be pushed away instead.

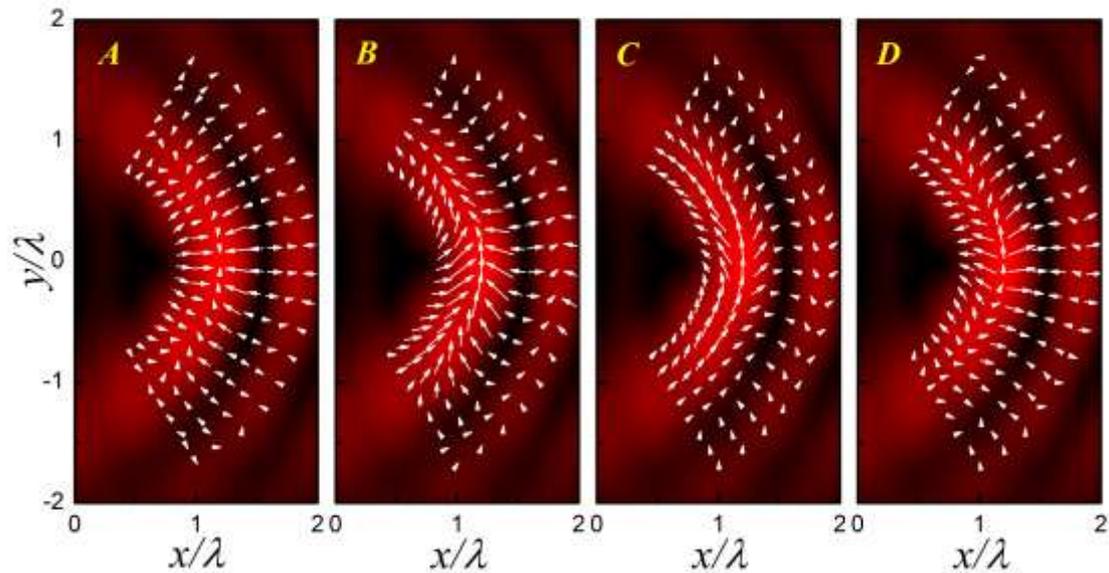

**Figure 6 | ARF distributions for air bubbles with different $kR$.** The corresponding bubble sizes are marked in Fig. 5(a). The maximal magnitudes of the dimensionless ARFs for the four cases are 531.9, 7756.9, 7511.6 and 1154.8, respectively.

To demonstrate the $kR$ dependent bubble manipulation more clearly, in Fig. 6 we present the ARF distributions for the four cases marked in Fig. 5(a), where the cases *A* and *B* are before resonance and *C* and *D* after resonance. Again, the pressure field is provided for comparison. For the case *A* far away from the resonance, as anticipated from Eq. (2) and Fig. 4, the soft bubble tends to be attracted into the bright pressure stripes. The angular scattering force is very weak; the air bubble in the main lobe is even trapped at the brightest spot by the gradient force. As $kR$ grows toward the monopole resonance, as manifested in the case *B*, the scattering force is greatly enhanced, which leads to a stable particle transport along the circular stripe. In contrast, for the case *C* slightly after resonance, the bubble tends to be repelled from the bright stripes, which leads to an unstable bending transport similar to the glycerin case. As $kR$ further increases to the case *D*, the scattering force reduces and the capability of bending transport becomes weaker.

Note that for the strong scattering case, the Eq. (2) fails to predict the stability in the radial direction, i.e., trapping in the bright or dark stripes. However, it is physically acceptable that the strong-field trapping preserves before resonance, according to the continuity from the low frequency limit. Once the resonance is crossed, a switch occurs and the weak-field trapping remains as $kR$ increases further. Therefore, the sound manipulation of air bubble can be flexibly switched by controlling the external frequency or bubble's size.

**Discussions**

The diffraction-free accelerating beam can keep its shape in propagation, even in a complex scattering environment. This self-healing property can be seen clearly in Fig. 7, where Fig. 7(a) exhibits the capability of bypassing a large and hard scatterer positioned in the central dark field region, and Figs. 7(b)-(d) demonstrate the beam recovery when impinging on the aforementioned three kinds of microparticles. This peculiar property endows the bending transport with a significant merit. That is, the proposed particle transport can not only bypass big obstacles, but also work in multiparticle scenarios. Similar phenomena can also be anticipated in the other 2D[33-35] and even 3D[39] self-accelerating beams, where the beam trajectories can be steered more flexibly.

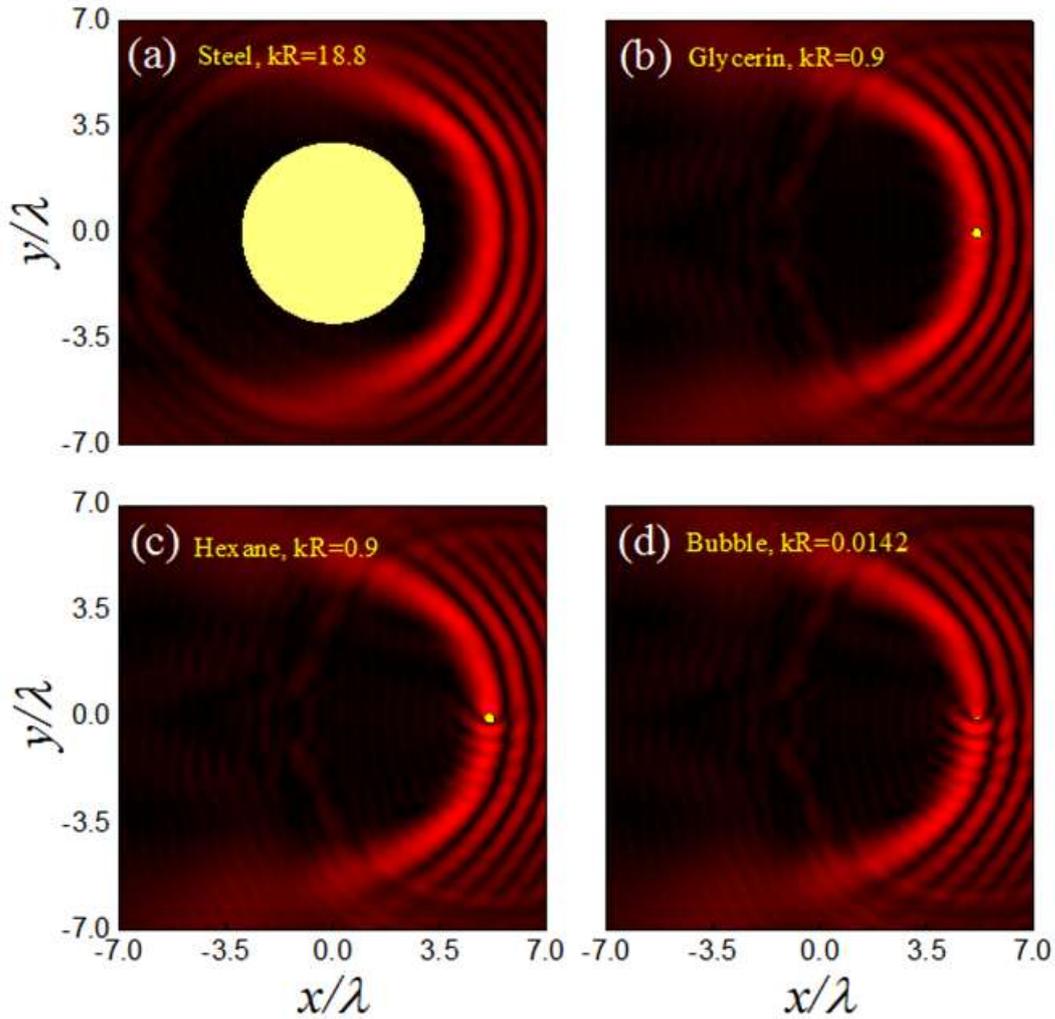

**Figure 7 | Self-healing property of the HB beam illustrated by** $\alpha = 30$. (a)-(d) Sound field patterns scattered by different objects (yellow circles), where the size parameters indicated in (b)-(d) correspond to the strongest scatterings in the aforementioned three systems.

This study is also instructive for its optic counterpart. For a dielectric microparticle, as the particle size increases the angular trapping can be tuned to the stable bending transport, since the subwavelength dielectric object prefers to stay in the strong-field region associated with considerable scattering force along the angular direction.[27,28,36] For a plasmonic particle, however, similar to the water-surrounded air bubbles discussed in the acoustic case, the manipulations become richer due to the subwavelength resonance of the localized surface plasmons.[40]

In conclusion, we have presented a comprehensive study on the microparticle manipulations through an acoustic HB beam. Strikingly different from the particle transport by conventional sound beams, which is usually along a straight line, the particle transport by the HB beam can be realized along a curved trajectory over a long distance. The bending angle is considerably large (nearly 180 degrees for large $\alpha$). It is worth pointing out that such self-accelerating beams can be practically realized by arranging a spatial array of transducers with elaborately-designed phase and amplitude profiles,[24,41,42] which enable the further experimental validation of the intriguing phenomena discussed here. Prospective applications can be anticipated for the sound manipulations proposed here, e.g., sorting, trapping and guiding microparticles.

**Method**

Here we give a brief introduction for evaluating the ARF acting on a spherical particle immersed in an ideal fluid, in which the viscous effect and thermal conductivity are neglected. The ARF can be calculated by integrating the time-averaged Brillouin radiation stress tensor $\langle \Pi \rangle$,

$$\mathbf{F} = -\oint_S \langle \Pi \rangle \cdot d\mathbf{A}, \tag{4}$$

where the differential area $d\mathbf{A}$ points to the outer normal of an arbitrary surface enclosing the particle. Specifically, the tensor $\langle \Pi \rangle$ can be written as

$$\langle \Pi \rangle = \left( \frac{|p|^2}{4\rho_0 c_0^2} - \frac{\rho_0 |\mathbf{u}|^2}{4} \right) \mathbf{I} + \frac{\rho_0 \mathrm{Re}(\mathbf{u}^*\mathbf{u})}{2}, \tag{5}$$

where $\mathbf{I}$ stands for a unit tensor and $\mathrm{Re}(\cdot)$ represents the real part.

To achieve the ARF for the particle located at $\mathbf{r}_p$, a critical step is to evaluate the self-consistent sound field of the scattering problem. For the convenience of the ARF calculation, we employ a spherical coordinate system defined from the particle center, i.e., $\mathbf{r}' = \mathbf{r} - \mathbf{r}_p = (r', \theta', \phi')$. In this coordinate system, the incident acoustic HB beam can be expanded as,

$$\psi_{HB} = \psi_0 \sum_{l=0}^{\infty} \sum_{m=-l}^{l} A_{lm} j_l(kr') Y_{lm}(\theta', \phi'), \tag{6}$$

where $j_l(\cdot)$ and $Y_{lm}(\cdot)$ are the spherical Bessel function and the spherical harmonic function, respectively. The incident coefficient

$$A_{lm} = \frac{1}{\psi_0 j_l(ka)} \int_{4\pi} \psi_{HB}(ka, \Omega') Y_{lm}^*(\theta', \phi') d\Omega' \tag{7}$$

can be solved numerically, where $a$ can be an arbitrary radius of the spherical region enclosing the particle (as long as $j_l(ka) \neq 0$), and $\Omega' = (\theta', \phi')$ is the solid angle in the spherical coordinate system. The scattering wave from the spherical particle can be written as

$$\psi_s = \sum_{l=0}^{\infty} \sum_{m=-l}^{l} B_{lm} h_l(kr') Y_{lm}(\theta', \phi'), \tag{8}$$

associated with scattering coefficient $B_{lm} = \sum_{l=0}^{\infty} t_l A_{lm}$, where $h_l(\cdot)$ is the spherical Hankel function of the first kind, and $t_l = \alpha_l + i\beta_l$ is the Mie scattering coefficient extracted from the boundary continuity at the particle surface.

Combining the incident and scattering waves, we can obtain the total field and consequently evaluate the ARF according to the surface integral in Eq. (4). After extending the integral surface into an infinite spherical surface,[23,38,43] the integral can be eventually expressed as a series related to the coefficients $A_{lm}$ and $t_l$, where the former depends on the property of the incident sound beam, and the latter characterizes the intrinsic property of the microparticle.

**Acknowledgments**

This work is supported by the National Natural Science Foundation of China (Grant Nos. 11174225, 11004155, 11374233, and J1210061); the National Basic Research Program of China (Grant No. 2015CB755500); and the Program for New Century Excellent Talents (NCET-11-0398).

**Author contributions**

C.Q. and Z.L. conceived the original idea and supervised the project. Y.L. performed the simulations with the help of S.X. and C.Q. Y.L. wrote the draft. C.Q., Z.L. and M.K. revised the manuscript. All authors contributed to scientific discussions of the manuscript.

**Competing financial interests**

The authors declare no competing financial interests.